\documentclass{ifacconf}
\usepackage{graphicx}      
\usepackage{natbib}        

\usepackage{amsmath}
\usepackage{amssymb}
\usepackage{amsfonts}
\usepackage{graphicx}
\usepackage{lipsum}
\usepackage{epstopdf}
\usepackage{algorithmic}
\usepackage{mathptmx}

\ifpdf
  \DeclareGraphicsExtensions{.eps,.pdf,.png,.jpg}
\else
  \DeclareGraphicsExtensions{.eps}
\fi
\DeclareMathAlphabet{\bit}{OML}{cmm}{b}{it}

\def\fH{\mathfrak{H}}

\def\<{\leqslant}           
\def\>{\geqslant}           

\def\d{\partial}
\def\wh{\widehat}

\def\Re{\mathrm{Re}}   
\def\Im{\mathrm{Im}}   

\def\cH{\mathcal{H}}   

\def\mR{\mathbb{R}}    

\def\Tr{\mathrm{Tr}}       
\def\rT{\mathrm{T}}        

\def\eps{\epsilon}

\def\rank{\mathrm{rank}}       

\def\rF{\mathrm{F}}        
\def\rHS{\mathrm{HS}}        
\def\bS{\mathbf{S}}         

\def\bE{\mathbf{E}}    
\def\bM{\mathbf{M}}    

\def\bP{\mathbf{P}}    

\def\[[[{[\![\![}   
\def\]]]{]\!]\!]}   

\def\bra{{\langle}}
\def\ket{{\rangle}}

\def\re{\mathrm{e}}        
\def\rd{\mathrm{d}}        

\def\cL{\mathcal{L}}

\def\bJ{\mathbf{J}}

\def\br{\mathbf{r}}
\def\x{\times}
\def\ox{\otimes}

\def\fF{\mathfrak{F}}

\def\fS{\mathfrak{S}}

\def\cF{\mathcal{F}}
\def\cK{\mathcal{K}}

\def\cD{\mathcal{D}}

\def\cN{\mathcal{N}}

\def\cG{\mathcal{G}}
\def\cI{\mathcal{I}}

\def\cQ{\mathcal{Q}}

\def\cB{\mathcal{B}}
\def\cE{\mathcal{E}}

\def\ups{\upsilon}

\begin{document}
\begin{frontmatter}

\title{Quantum Memory Optimisation Using Finite-Horizon, Decoherence  Time and Discounted Mean-Square Performance Criteria\thanksref{footnoteinfo}}%
\thanks[footnoteinfo]{This work is supported by the Australian Research Council grant DP240101494.}
\vspace{-2mm}
\author[First]{Igor G. Vladimirov},\quad
\author[First]{Ian R. Petersen},\quad
\author[Second]{Guodong Shi}

\vspace{-2mm}
\address[First]{School of Engineering, Australian National University, Canberra, ACT,
Australia (e-mail: igor.g.vladimirov@gmail.com, i.r.petersen@gmail.com)}
\address[Second]{Australian Centre for Robotics, University of Sydney, Camperdown, Sydney, NSW, Australia (e-mail: guodong.shi@sydney.edu.au)}

\begin{abstract}                
This paper is concerned with open quantum memory systems for approximately retaining quantum information, such as initial dynamic variables or quantum states to be stored over a bounded time interval. In the Heisenberg picture of quantum dynamics,  the deviation  of the system variables from their initial values lends itself to closed-form computation in terms of tractable moment dynamics for open quantum harmonic oscillators
and finite-level quantum systems governed by linear or quasi-linear Hudson-Parthasarathy quantum stochastic differential equations,
respectively.
This tractability is used in a recently proposed   optimality criterion for varying the system parameters so as to  maximise the memory decoherence time when the mean-square deviation
achieves a given critical threshold. The memory decoherence time maximisation  approach is extended beyond the previously considered low-threshold
asymptotic approximation and to Schr\"{o}dinger type mean-square deviation functionals for the reduced system state governed by the Lindblad master equation.  We link this approach with the minimisation of the mean-square deviation functionals at a finite time horizon and with their discounted version which quantifies the averaged performance of the quantum system as a temporary memory under a Poisson flow of storage requests.
\vspace{-3mm}
\end{abstract}

\vspace{-2mm}
\end{frontmatter}
\section{Introduction}
\vspace{-2mm}

Quantum information processing is an inherent part of quantum computation and quantum communication architectures [\cite{NC_2000}]. Its storage  relies on the engineered ability of quantum systems to approximately retain quantum states or dynamic variables over a sufficiently long period of time. This task is performed ideally  by an isolated quantum system with zero Hamiltonian, in which case, the system variables and the system state remain constant over the course of time. In reality, interaction of the system with the environment, including external fields, leads to open quantum dynamics.

In quantum stochastic calculus  [\cite{HP_1984,P_1992}],  the Heisenberg picture  evolution  of open quantum systems is modelled   by quantum stochastic differential equations (QSDEs) driven by a quantum Wiener process which represents the environmental noise. The type of the resulting QSDEs depends on the algebraic or commutation structure of the system variables and the dependence of the system Hamiltonian and the system-field coupling operators on them.

In particular, open quantum harmonic oscillators (OQHOs) with position-momentum variables and finite-level open quantum systems with Pauli matrix [\cite{S_1994}] type variables  are governed by linear and quasi-linear QSDEs, respectively. In both cases, and with the external input fields in the vacuum state, the system variables have tractable moment dynamics, which allows for closed-form computation of their mean-square deviation from the initial conditions.

The moment  tractability is used in a recently proposed   optimality criterion [\cite{VP_2024_ANZCC}] for varying the system parameters so as to  maximise the memory decoherence time at which the mean-square deviation functional achieves a given critical threshold specified by a dimensionless fidelity level and a reference scale. The resulting quantum memory optimisation problems have been solved for interconnections of OQHOs and finite-level systems through  direct  energy and  field-mediated coupling in the low-threshold (that is, high-fidelity  or short-horizon) asymptotic approximation [\cite{VP_2024_EJC,VP_2025_ANZCC,VPS_2025_CDC}].

The present paper extends the memory decoherence time maximisation  approach beyond the short-horizon approximation  and towards Schr\"{o}dinger picture type mean-square deviation functionals involving the Hilbert-Schmidt distance between the reduced system state (governed by the Lindblad master equation) and the initial state.  Using the first and second-order conditions of optimality,   we relate this approach to the minimisation of the mean-square deviation functionals of both types at a given finite time horizon. Also, we discuss an exponentially discounted version of these functionals which describes the averaged performance of the quantum system as a temporary memory under a classical Poisson process of storage requests (adopted from queueing theory).

The paper is organised as follows.
Section~\ref{sec:sys} specifies the class of open quantum systems under consideration.
Section~\ref{sec:MSDF} describes the Heisenberg and Schr\"{o}dinger type mean-square deviation functionals for dynamic variables and reduced quantum states.
Section~\ref{sec:long} relates the memory decoherence time maximisation to the minimisation of the deviation functionals at a given time horizon.
Section~\ref{app:disc} discusses the discounted versions of these functionals and their computation for OQHOs and finite-level quantum systems.
Section~\ref{sec:conc} provides concluding remarks.

\section{Open quantum stochastic systems}\label{sec:sys}

We consider an open quantum stochastic system with dynamic variables $X_1(t), \ldots, X_n(t)$ which are time-varying self-adjoint operators (the time argument $t\> 0$ is often omitted)  on the tensor-product system-field space
\begin{equation}
\label{fH}
    \fH := \fH_0 \ox \fF.
\end{equation}
Here, $\fH_0$ is the initial system space (a domain for the initial system variables $X_1(0), \ldots, X_n(0)$),   and $\fF$ is a symmetric Fock space [\cite{PS_1972}] for the action of an even number of quantum Wiener processes $W_1, \ldots, W_m$ which are time-varying self-adjoint operators modelling  the external bosonic fields.  The Heisenberg  picture evolution of the column-vector $X:= (X_k)_{1\< k\< n}$ is driven by  $W:= (W_k)_{1\< k \< m}$ according to a Markovian Hudson-Parthasarathy QSDE [\cite{HP_1984,P_1992}] with the identity scattering matrix (thus eliminating from consideration the gauge processes associated with the photon exchange between the fields):
\begin{equation}
\label{dX}
    \rd X = \cG(X)\rd t + \cB\rd W,
    \qquad
    \cB:= -i[X,L^\rT].
\end{equation}
Here, $i:= \sqrt{-1}$ is the imaginary unit, $(\cdot)^\rT$ is the usual transpose,   and \begin{equation}
\label{XLcomm}
    [X,L^\rT]
    =
    ([X_j,L_k])_{1\< j \< n, 1\< k \< m} = - [L,X^\rT]^\rT
\end{equation}
is the matrix of commutators $[\alpha, \beta]:= \alpha \beta - \beta \alpha$ between the system variables and the system-field coupling operators $L_1, \ldots, L_m$. The latter are time-varying self-adjoint operators which acquire dependence on time as functions (for example, polynomials with constant coefficients) of $X_1, \ldots, X_n$ and are assembled into a vector $L:= (L_k)_{1\< k \< m}$. Accordingly, $\cB$ in (\ref{dX}) is an $(n\x m)$-matrix whose entries are self-adjoint operator-valued functions of $X_1, \ldots, X_n$.
Also,
$\cG$ is the Gorini-Kossakowski-Sudarshan-Lindblad (GKSL) superoperator [\cite{GKS_1976,L_1976}] acting on a system operator $\zeta$ (a function of the system variables $X_1, \ldots, X_n$) on the space $\fH$   as
\begin{equation}
\label{cG}
   \cG(\zeta)
   :=
   i[H,\zeta] +
   \cD(\zeta),
\end{equation}
where $H$ is the system Hamiltonian which is a self-adjoint operator on $\fH$ organised as a function of the system variables.
The GKSL superoperator $\cG$  involves the decoherence superoperator    $\cD$ of the form
\begin{equation}
\label{cD}
    \cD(\zeta)
    :=
    \frac{1}{2}
    (
        [L^{\rT},\zeta]\Omega L  + L^{\rT}\Omega [\zeta,L]
    ),
\end{equation}
where $\Omega$ is a complex positive semi-definite Hermitian matrix of order $m$ given by
\begin{equation}
\label{Omega}
    \Omega
    :=I_m + iJ = \Omega^* \succcurlyeq 0,
    \quad
        J :=
    I_{m/2} \ox \bJ,
    \quad
    \bJ:=
    \begin{bmatrix}
        0 & 1\\
        -1 & 0
    \end{bmatrix},
\end{equation}
with $(\cdot)^*:= {\overline{(\cdot)}}^\rT$ the complex conjugate transpose, and $I_m$ the identity matrix of order $m$. The matrix $\Omega$ is the quantum Ito matrix for the future-pointing increments of the quantum Wiener processes $W_1, \ldots, W_m$:
\begin{equation}
\label{dWdW}
    \rd W\rd W^{\rT} = \Omega \rd t.
\end{equation}
Its imaginary part $J = \Im \Omega$ in (\ref{Omega}) is a real antisymmetric matrix of order $m$ specifying the canonical commutation relations (CCRs) $[\rd W, \rd W^{\rT}] = 2iJ\rd t$ and the two-point CCRs $[W(s), W(t)^\rT] = 2i \min(s,t) J$ for all $s,  t \> 0$. In (\ref{dX}), the GKSL superoperator  $\cG$, given by (\ref{cG}), (\ref{cD}),  is applied to the vector $X$ in an entrywise fashion, so that
\begin{equation}
\label{cGvec}
    \cG(X) = i [H, X]
    -
    \frac{1}{2}
    (
        [X,L^{\rT}]\Omega L  + (L^{\rT}\Omega [L,X^\rT])^\rT
    )
\end{equation}
in view of the antisymmetry of the commutator as in (\ref{XLcomm}).
At any time $t\> 0$, the system variables $X_1(t), \ldots, X_n(t)$ act on the subspace
\begin{equation}
\label{fHt}
    \fH_t:= \fH_0\ox \fF_t
\end{equation}
of (\ref{fH}),
where $\{\fF_t\}_{t \> 0}$ is an increasing family of subspaces of $\fF$ which form the Fock space filtration. Accordingly, the adaptedness of quantum processes under consideration is understood with respect to the system-field space filtration $\{\fH_t\}_{t\> 0}$ given by (\ref{fHt}).
The QSDE (\ref{dX}), specified by (\ref{cG}), (\ref{cD}), generates the quantum stochastic flow
\begin{equation}
\label{uni}
    X(t)
    =
    (U(t)^{\dagger} (X_k(0)\ox \cI_\fF) U(t))_{1\< k \< n},
\end{equation}
where $(\cdot)^{\dagger}$ is the operator adjoint,  $\cI_\fF$ is the identity operator on $\fF$ (so that the operation $\cdot \ox \cI_\fF$ extends operators from $\fH_0$ to $\fH$),
and $U(t)$ is a unitary operator on $\fH$ governed by a stochastic Schr\"{o}dinger equation
\begin{align*}
    \rd U(t)
    & =
    -\Big(i\big(H_0\rd t + L_0^{\rT} \rd W(t)\big) + \frac{1}{2}L_0^{\rT}\Omega L_0\rd t\Big)U(t)\\
    &= -U(t)
    \Big(i\big(H\rd t + L^{\rT} \rd W(t)\big) + \frac{1}{2}L^{\rT}\Omega L\rd t\Big),
\end{align*}
with the initial condition $U(0) = \cI_\fH$, where $H_0$, $L_0$ are the initial Hamiltonian and the vector of the initial system-field coupling operators acting on $\fH_0$. Note that the dependence of $H$, $L$ on $X$ is the same as that of $H_0$, $L_0$ on $X_0:= X(0)$ and is thus preserved over the course of time. 
It is assumed that the input fields are in the vacuum state [\cite{P_1992}] $\ups$ on the Fock space $\fF$, and   the system-field quantum state is given by a tensor-product density operator
\begin{equation}
\label{rho}
    \rho := \sigma_0 \ox \ups
\end{equation}
on $\fH$ in (\ref{fH}),
where $\sigma_0$ is the initial system state on $\fH_0$.
The reduced system state
\begin{equation}
\label{sig}
    \sigma(t):= \Tr_\fF (U(t) \rho U(t)^\dagger)
\end{equation}
(with $\Tr_\fF(\cdot)$ the partial trace of an operator on $\fH$ over the Fock space $\fF$, tracing out the  field variables)  is a time-varying density operator on the initial system space $\fH_0$ satisfying the Lindblad master equation
\begin{equation}
\label{sigdot}
    \dot{\sigma}
    =
    -
    i[H_0,\sigma]
    +
    (L_0\sigma)^\rT \overline{\Omega} L_0 - \frac{1}{2}\{L_0^\rT \Omega L_0, \sigma\}
    =:
    \cL(\sigma)
\end{equation}
with the initial condition $\sigma(0) = \sigma_0$ from (\ref{rho}). Here, $\dot{(\ )}$ is the time derivative, $\{\alpha, \beta\}:= \alpha\beta + \beta \alpha$ is the anticommutator of operators, and $\overline{\Omega} = \Omega^\rT = I_m - iJ$ in view  of (\ref{Omega}). The superoperator   $\cL$ maps operators on $\fH_0$ to those with zero trace and preserves self-adjointness. The reduced system state (\ref{sig}) allows the quantum expectations of functions of the system variables and their evolution  to be represented in the Schr\"{o}dinger picture as
\begin{align}
\nonumber
    \bE f(X(t))
    & = \Tr(\rho f(X(t)))
    =
    \Tr (\rho U(t)^{\dagger} (f(X_0)\ox \cI_\fF) U(t))    \\
\label{EfX}
    & =
    \Tr (\sigma(t) f(X_0))
\end{align}
in view of (\ref{uni}).
Therefore, the one-point moments of the system variables at any time $t\> 0$  can be computed by finding the reduced system state $\sigma(t)$ from the Lindblad master equation (\ref{sigdot}) instead of $X(t)$ from the QSDE (\ref{dX}). However, the latter allows for computation of multi-point moments  of the system variables at different times, including the two-point second-moment matrix
\begin{equation}
\label{EXX}
    S(u,v):= \bE(X(u) X(v)^\rT),
    \qquad
    u,v\> 0.
\end{equation}
This matrix is involved in the one-point second-moment matrix
$$
    \bE (\xi(t)\xi(t)^\rT) =  S(0,0) + S(t,t) - S(t,0)-S(0,t)
$$
for the deviation
\begin{equation}
\label{xi}
    \xi(t):= X(t)-X_0
\end{equation}
of the system variables at time $t\>0$ from their initial values (with $\xi(0)=0$).

\section{Mean-square deviation functionals}\label{sec:MSDF}

As a quantum memory, the system under consideration can be oriented to an approximate preservation of $s\< n$ linear combinations (in particular, a subset) of the system variables captured in the quantum process
\begin{equation}
\label{FX}
    \varphi(t):= FX(t),
\end{equation}
where $F\in \mR^{s\x n}$ is a given full row rank matrix.  The memory performance can be quantified in terms of the mean-square deviation functional [\cite{VP_2024_ANZCC,VP_2024_EJC}]
\begin{equation}
\label{Del}
    \Delta(t) := \bE (\eta(t)^\rT\eta(t)) = \bE (\xi(t)^\rT \Sigma \xi(t)),
\end{equation}
where
\begin{equation}
\label{eta}
    \eta(t):= \varphi(t) - \varphi(0) = F \xi(t),
\end{equation}
in accordance with (\ref{xi}),
and
\begin{equation}
\label{FF}
    \Sigma := F^\rT F
\end{equation}
is a real positive semi-definite symmetric matrix with $\rank \Sigma = s$.

\begin{lem}
\label{lem:Deldots}
The quantity (\ref{Del}) is a smooth function of time satisfying
$\Delta(0) = 0$, and its first two derivatives at any time $t\> 0$ are computed as
\begin{align}
\nonumber
    \dot{\Delta}
    & = 2 \Re \bE (\xi^\rT \Sigma \cG(X)) + \bra \Sigma, \bE ([X,L^\rT] \Omega [L, X^\rT])\ket_\rF\\
\label{Deldot}
    & = \bE \cG(X^\rT \Sigma X) - 2\Re \bE (X_0^\rT \Sigma \cG(X)), \\
\label{Delddot}
    \ddot{\Delta}
    & = \bE \cG^2(X^\rT \Sigma X) - 2\Re \bE (X_0^\rT \Sigma \cG^2(X)),
\end{align}
where $\cG^2:= \cG \circ \cG$ is the composition  of the GKSL superoperator $\cG$ from (\ref{cG}), (\ref{cGvec}) with itself,  and
$\bra \cdot, \cdot\ket_\rF$ is the Frobenius inner product of complex or real matrices (which will also be formally applied when one of the matrices consists of operators).
\hfill$\square$
\end{lem}
\begin{pf}
Since $\rd \xi(t) = \rd X(t)$ in view of (\ref{xi}),
application of the quantum Ito lemma yields
\begin{align}
\nonumber
    \rd (\xi^\rT \Sigma \xi)
    = &
    \rd X^\rT \Sigma \xi + \xi^\rT \Sigma \rd X + \rd X^\rT \Sigma \rd X\\
\nonumber
     = &
    (\cG(X)^\rT \Sigma \xi + \xi^\rT \Sigma \cG(X) + \bra \Sigma, \cB \Omega \cB^\rT\ket_\rF)\rd t\\
\nonumber
    & +
    \rd W^\rT \cB^\rT \Sigma \xi
    +
    \xi^\rT \Sigma  \cB \rd W\\
\nonumber
     = &
    (2 \Re (\cG(X)^\rT \Sigma \xi) + \bra \Sigma, \cB \Omega \cB^\rT\ket_\rF)\rd t\\
\label{detaeta}
    & +
    2 \Re (\xi^\rT \Sigma  \cB ) \rd W.
\end{align}
Here, use is made of the QSDE (\ref{dX}) along with the commutativity $[\rd W(t), \zeta(s)] = 0$ for any adapted quantum process $\zeta$ at any times $t \> s \> 0$, the quantum Ito relation
\begin{equation}
\label{dXdX}
    \rd X \rd X^\rT
    =
    \cB \rd W \rd W^\rT \cB^\rT
    =
    \cB \Omega \cB^\rT\rd t
\end{equation}
from (\ref{dWdW}),
the symmetry of the matrix (\ref{FF}),  and the extension $\Re z  := \frac{1}{2}(z+ z^\#)$ of the real part to matrices $z$  of operators, with $(\cdot)^\#$ the entrywise operator adjoint. Since the extended operation  $\Re(\cdot)$  commutes with the quantum expectation $\bE(\cdot)$  (in the sense that $\bE \Re z = \Re \bE z$), then by taking the latter on both sides of (\ref{detaeta})
and using the fact that the quantum diffusion term makes no contribution to this expectation due to the external field $W$ being  in the vacuum state, the time derivative of (\ref{Del}) takes the form
\begin{align}
\nonumber
    \dot{\Delta}
    & =
    \bE (2 \Re (\cG(X)^\rT \Sigma \xi) + \bra \Sigma, \cB \Omega \cB^\rT\ket_\rF)\\
\label{Deldot1}
    & =
    2\Re \bE (\cG(X)^\rT \Sigma \xi) + \bra \Sigma, \bE (\cB \Omega \cB^\rT)\ket_\rF.
\end{align}
In view of the second equalities in (\ref{dX}),  (\ref{XLcomm}), the quantum Ito matrix of the system variables in (\ref{dXdX}) can be represented as $\cB \Omega \cB^\rT = [X,L^\rT] \Omega [L,X^\rT]$,  and hence, (\ref{Deldot1}) leads to the first equality in (\ref{Deldot}). The second equality in (\ref{Deldot}) is obtained in a similar fashion from
an alternative (yet equivalent) representation of (\ref{detaeta}):
\begin{align*}
    \rd (\xi^\rT \Sigma \xi)
    = &
    \rd (X^\rT \Sigma X) - 2\Re (X_0^\rT \Sigma \rd X)\\
     = &
    (\cG(X^\rT \Sigma X)  - 2\Re (X_0^\rT \Sigma \cG(X))\rd t\\
    & - (i [X^\rT \Sigma X, L^\rT]
    + 2\Re (X_0^\rT \Sigma \cB)) \rd W\\
     = &
    (\cG(X^\rT \Sigma X)  - 2\Re (X_0^\rT \Sigma \cG(X))\rd t
     +
    2 \Re (\xi^\rT \Sigma  \cB ) \rd W,
\end{align*}
which follows from the relation $\xi^\rT \Sigma \xi = X^\rT \Sigma X - 2\Re (X_0^\rT \Sigma X) + X_0^\rT \Sigma X_0$ and the derivation property of the commutator whereby  $[X^\rT \Sigma X, L^\rT] = X^\rT \Sigma [X, L^\rT] - ([L,X^\rT]\Sigma X)^\rT$. Repeated application of this technique to the second equality in (\ref{Deldot}) establishes (\ref{Delddot}).
 \hfill$\blacksquare$
\end{pf}

The right-hand side of (\ref{Deldot}) involves not only $X(t)$ but also the initial condition  $X_0$. Therefore, its computation cannot  be carried out in terms of the reduced system state $\sigma(t)$, thus again indicating the insufficiency of the latter for computing the Heisenberg picture type mean-square deviation functional (\ref{Del}).

On the other hand,  for Schr\"{o}dinger picture formulations of quantum memory performance,  alternative functionals can be  considered which quantify the deviation of $\sigma(t)$ from $\sigma_0$. One of them is provided by the squared Hilbert-Schmidt distance:
\begin{equation}
\label{DelHS}
    \Gamma(t)
    :=
    \|\gamma(t)\|_\rHS^2 = \Tr (\gamma(t)^2),
    \qquad
    \gamma(t):= \sigma(t)-\sigma_0,
\end{equation}
where the self-adjointness of $\gamma$ (inherited from the density operator $\sigma$)   is taken into account. In comparison with (\ref{Deldot}), (\ref{Delddot}),  the time derivatives of the functional $\Gamma$ lend themselves to a more straightforward calculation as
$$
    \dot{\Gamma}
     =
    2 \bra \gamma, \cL(\sigma)\ket_\rHS,
    \quad
    \ddot{\Gamma}
    =
    2 \bra \gamma, \cL^2(\sigma)\ket_\rHS
    +
    2\|\cL(\sigma)\|_\rHS^2
$$
by using (\ref{sigdot}), with
$\bra\cdot, \cdot \ket_\rHS$ the Hilbert-Schmidt inner product of operators on $\fH_0$. While, as mentioned above,  the knowledge of the reduced system state $\sigma$  is insufficient for computing $\Delta$ in (\ref{Del}), the functional $\Gamma$ in (\ref{DelHS}) leads to an upper bound for the deviation of the one-point moments of the system variables in (\ref{EfX}) from their initial values:
\begin{align*}
    |\bE f(X(t)) - \bE f(X_0)|
    & =
    |\bra \gamma(t), f(X_0)\ket_\rHS |
    \<
    \|f(X_0)\|_\rHS
    \sqrt{\Gamma(t)}
\end{align*}
for     any $t\> 0$,
where use is made of the Cauchy-Bunyakovsky-Schwarz inequality.

Despite the different definitions (\ref{Del}), (\ref{DelHS}), both mean-square deviation functionals  $\Delta$ and $\Gamma$ are nonnegative  smooth functions of time with zero initial conditions. If  their values over a bounded time interval $[0,T]$ (for a given horizon $T>0$) are relatively small, that is, so is at least one of the corresponding largest values
\begin{equation}
\label{Delmax}
    \Delta_T
    :=
    \max_{0\< t \< T}
    \Delta(t) ,
    \qquad
    \Gamma_T
    :=
    \max_{0\< t \< T}
    \Gamma(t),
\end{equation}
then this indicates the ability of the system to approximately retain its initial dynamic variables of interest specified by  $\varphi$ in (\ref{FX}) or the reduced quantum state $\sigma$ from   (\ref{sig}) over the interval.

\section{DECOHERENCE TIME MAXIMISATION BEYOND SHORT HORIZONS}\label{sec:long}

Concerning the mean-square deviation functional $\Delta$ from (\ref{Del}) for concreteness
(the subsequent discussion is equally applicable to $\Gamma$ in (\ref{DelHS})),
the requirement of smallness of $\Delta_T$ in (\ref{Delmax}) at a given time horizon $T$ admits an alternative formulation  in terms of prolonging the time while $\Delta$ remains below a given level. More precisely, this is quantified by the
memory decoherence time [\cite{VP_2024_ANZCC,VP_2024_EJC}]
\begin{align}
\nonumber
    \tau(\eps)
    & :=
    \sup
    \{
        T >0:\
        \Delta_T < \eps\Delta_*
    \}\\
\label{tau}
    & =
    \min
    \{
        t\> 0:\
        \Delta(t)
        \>
        \eps
        \Delta_*
    \},
\end{align}
that is, the first moment of time at which the nondecreasing function $0\< T\mapsto \Delta_T \in \mR_+$ (with $\Delta_0 = 0$) in (\ref{Delmax}) achieves a critical threshold value $\eps \Delta_*$.
Here, $\Delta_*>0$ is a given  reference  scale, and $\eps >0 $ is a dimensionless fidelity parameter. Note that
\begin{equation}
\label{Deleps}
    \Delta(t)
    \<
    \Delta_t
    <
    \Delta(\tau(\eps))
    =
    \Delta_{\tau(\eps)}
    =
    \eps\Delta_*
    \quad
    {\rm for\ all}\
    t \in
    [0, \tau(\eps))
\end{equation}
in view of (\ref{Delmax}), (\ref{tau}) and the Intermediate Value Theorem applied to the continuous function $\Delta$. Also, $\tau(\eps)$ is a nondecreasing function, which satisfies $\tau(\eps)>0$ for any $\eps>0$ and extends to $\tau(0):= \tau(0+) = \lim_{\eps\to 0+} \tau(\eps) = 0$ by continuity. Accordingly,  small values of $\eps$ correspond to the high-fidelity (or short-horizon) limit. With $\Delta$ being the mean-square deviation functional (\ref{Del}) in view of (\ref{eta}), a natural reference scale, associated with the initial system conditions to be stored,  is provided by
\begin{equation}
\label{Del*}
    \Delta_*
    :=
    \bE (\varphi(0)^\rT \varphi(0) )
    =
    \bra \Sigma, S(0,0)\ket_\rF = \|F\sqrt{P}\|_\rF^2, \!
\end{equation}
where
\begin{equation}
\label{P}
  P := \Re S(0,0)
\end{equation}
is a real positive semi-definite symmetric matrix coming from (\ref{EXX}). Here,
$\|\cdot\|_\rF$ is the Frobenius norm of matrices, and  use is made of (\ref{FX}), (\ref{FF}) along with the fact that the matrix $\Im S(0,0)$ is antisymmetric and thus orthogonal to $\Sigma$ (that is, $\bra \Sigma, \Im S(0,0)\ket_\rF = 0$). The trivial case of $\Delta_*=0$ in (\ref{Del*}) is eliminated from consideration.

The above constructs (\ref{tau})--(\ref{Del*})  can be adapted to the Schr\"{o}\-din\-ger picture mean-square deviation functional  $\Gamma$ in (\ref{DelHS}) instead of $\Delta$ in an almost verbatim fashion, by replacing $\Delta$ with $\Gamma$,   except that an appropriate counterpart of $\Delta_*$ in (\ref{Del*}) is provided by $\Gamma_*:= \|\sigma_0\|_\rHS^2$.
Moreover, in what follows, we will mainly use the basic properties of $\Delta(t)$ as a function of time $t\> 0$,  such as smoothness, nonnegativity and zero initial value, rather than its specific structure of a particular mean-square deviation functional.

\begin{lem}
\label{lem:nonneg}
For any $\eps\> 0$, the time derivative of the function $\Delta$ from   (\ref{Del}) at the decoherence time $\tau(\eps)$ in  (\ref{tau}) is nonnegative:
\begin{equation}
\label{Deltau}
  \dot{\Delta}(\tau(\eps)) \> 0 .
\end{equation}
\hfill$\square$
\end{lem}
\begin{pf}
In the limiting case of $\eps=0$, the inequality (\ref{Deltau}) takes the form \begin{equation}
\label{Deldot0}
    \dot{\Delta}(0)\> 0
\end{equation}
(with the right derivative  at $0$) and follows from the fact that $\Delta(0)=0 \< \Delta(t)$ for any $t\> 0$. Now, consider a fixed but otherwise arbitrary $\eps >0$.
In order to prove (\ref{Deltau}) by contradiction, assume that it is violated: $\dot{\Delta}(\tau(\eps)) < 0$. Then there exists $t\in (0, \tau(\eps))$, sufficiently close to $\tau(\eps)$,  such that $\Delta(t) > \Delta(\tau(\eps)) = \eps \Delta_*$. However, the latter inequality is impossible in view of the inequalities in (\ref{Deleps}). This contradiction establishes (\ref{Deltau}).
\hfill$\blacksquare$
\end{pf}

In view of Lemma~\ref{lem:nonneg},  we will be particularly concerned with those values of $\eps\> 0$, for which  (\ref{Deltau}) holds as a strict inequality:
\begin{equation}
\label{Deldotpos}
  \dot{\Delta}(\tau(\eps)) >0.
\end{equation}
Such values will be referred to as \emph{regular} and their set is denoted by
\begin{equation}
\label{cE}
  \cE
  :=
  \{
        \eps \>0:\
        \dot{\Delta}(\tau(\eps)) >0
  \}.
\end{equation}
Its complement  $\mR_+ \setminus \cE =   \{
        \eps \>0:
        \dot{\Delta}(\tau(\eps)) =0
  \} \subset \frac{1}{\Delta_*} \Delta(\{t\> 0: \dot{\Delta}(t) = 0\})$  has zero Lebesgue measure by the Morse-Sard Theorem [\cite{M_1939,S_1942}].
Furthermore, the set $\cE$ in (\ref{cE}) contains a neighbourhood of  $0$ in $\mR_+$ whenever (\ref{Deldot0}) holds as a strict inequality:
\begin{equation}
\label{Deldotpos0}
    \dot{\Delta}(0)>0.
\end{equation}
In this case, the decoherence time (\ref{tau}) admits a quadratically truncated  Taylor series expansion
\begin{equation}
\label{tauquad}
    \tau(\eps)
    =
    \underbrace{\tau'(0)\eps + \frac{1}{2}\tau''(0)\eps^2}_{\wh{\tau}(\eps)} + o(\eps^2),
    \qquad
    {\rm as}\
    \eps \to 0+,
\end{equation}
where the first two right derivatives of $\tau$ at $\eps=0$ are computed as
\begin{equation}
\label{tau'''0}
    \tau'(0) = \frac{\Delta_*}{\dot{\Delta}(0)},
    \qquad
    \tau''(0)
    =
    -
    \frac{ \Delta_*^2\ddot{\Delta}(0)}
    {\dot{\Delta}(0)^3}
\end{equation}
in terms of the  right time derivatives of $\Delta$ from  (\ref{Deldot}), (\ref{Delddot}) at $t=0$. The asymptotic behaviour (\ref{tauquad}) of $\tau$ in the high-fidelity (or, equivalently, short-horizon) limit was used in [\cite{VP_2024_ANZCC,VP_2024_EJC}] in the problem of  maximising the approximate memory decoherence time $\wh{\tau}(\eps) \to \sup$ (at a fixed small $\eps>0$) over system parameters. In order to extend this optimisation approach beyond short horizons, that is, to arbitrary values of the fidelity parameter $\eps>0$,  the following lemma provides an additional insight into the structure of regular fidelity levels.

\begin{lem}
\label{lem:tau'}
The set (\ref{cE}) of regular values of the fidelity parameter is open, the memory decoherence time (\ref{tau}) on it inherits smoothness from $\Delta$, and
\begin{equation}
\label{tau'}
  \tau'(\eps)
  =
  \frac{\Delta_*}{\dot{\Delta}(\tau(\eps))},
  \quad
    \tau''(\eps)
      =
    -
    \frac{ \Delta_*^2\ddot{\Delta}(\tau(\eps))}
    {\dot{\Delta}(\tau(\eps))^3},
  \qquad
  \eps \in \cE,
\end{equation}
where $\Delta_*$ is given by (\ref{Del*}), and $\dot{\Delta}$, $\ddot{\Delta}$ are found in (\ref{Deldot}), (\ref{Delddot}). \hfill$\square$
\end{lem}
\begin{pf}
Let $\eps_0>0$ be a regular value of the fidelity parameter from (\ref{cE}), so that $\dot{\Delta}(T)>0$, where $T:= \tau(\eps_0)$.  Then there exists a $\delta \in (0, T)$ such that $\dot{\Delta}$ is also positive in the $\delta$-neighbourhood of $T$:
\begin{equation}
\label{DeldotposV}
    \dot{\Delta}(t)>0,
    \qquad
    t \in (T-\delta, T+\delta) =: V.
\end{equation}
Hence, the restriction $\Delta|_V: V \to \Delta(V)$ is a strictly increasing function which has a strictly increasing inverse $(\Delta|_V)^{-1}: \Delta(V)\to V$ on the image interval $\Delta(V) = (\Delta(T-\delta), \Delta(T+\delta))$, which is also continuously differentiable by the Inverse Function Theorem.  Since, by a combination of (\ref{Deleps}) with (\ref{DeldotposV}),
\begin{equation}
\label{<<}
    \Delta(T-\delta) \< \Delta_{T-\delta} < \eps_0 \Delta_* < \Delta(T+\delta),
\end{equation}
then a smaller neighbourhood $N:= (\Delta_{T-\delta},\Delta(t+\delta)) \subset \Delta(V)$ of $\eps_0\Delta_*$ (see Fig.~\ref{fig:msdf})
\begin{figure}
\begin{center}
\includegraphics[width=8.75cm]{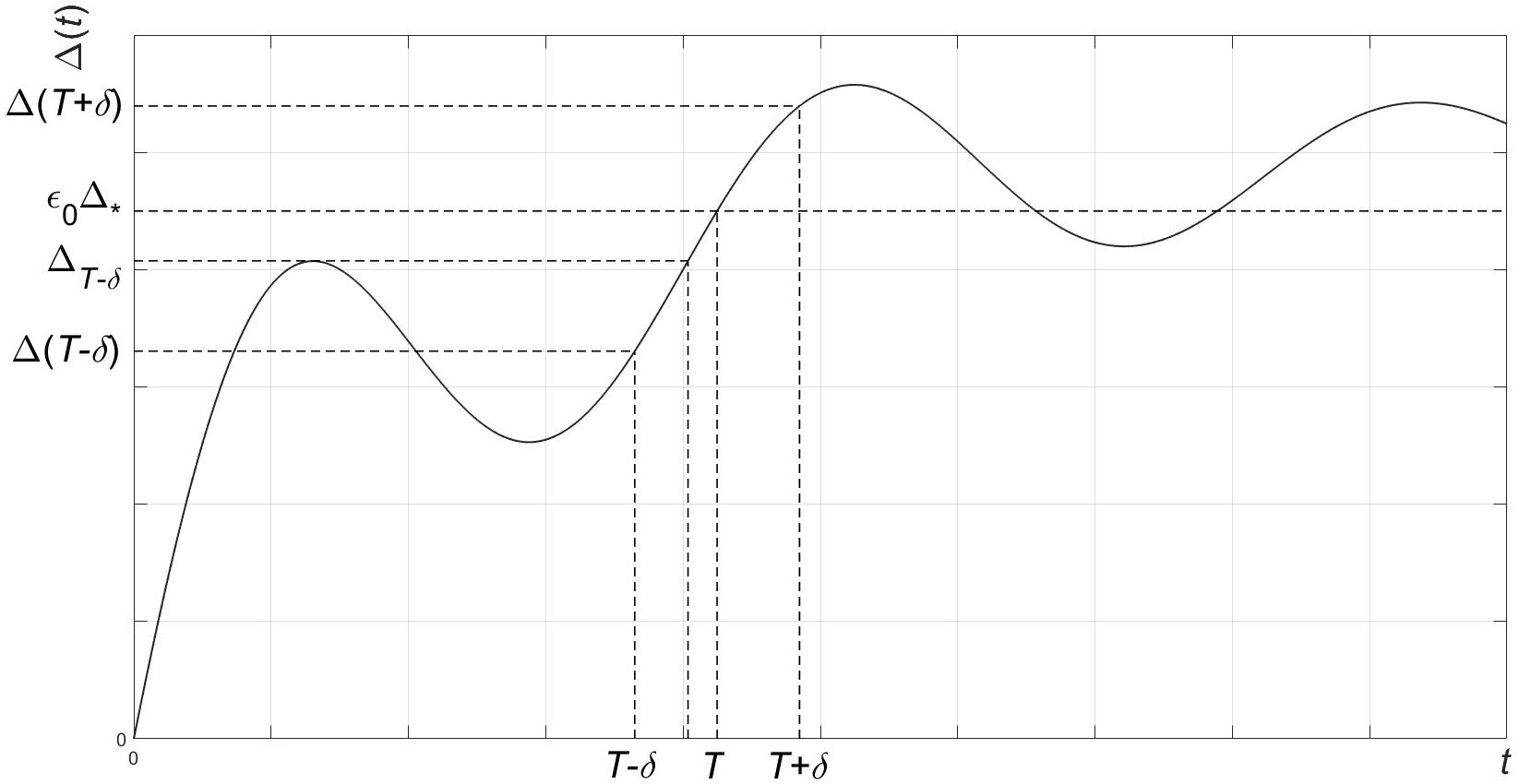}    
\caption{An illustration of the local inverse of the function $\Delta$ in a neighbourhood of $T:=\tau(\eps_0)$ with a regular fidelity level $\eps_0$ in the sense of (\ref{Deldotpos}).}
\label{fig:msdf}
\end{center}
\end{figure}
allows the decoherence time (\ref{tau}) to be related to the local inverse of $\Delta$ as
\begin{equation}
\label{locinv}
    \tau(\eps)
    =
    (\Delta|_V)^{-1}(\eps\Delta_*)
    \in ((\Delta|_V)^{-1}(\Delta_{T-\delta}), T+\delta) \subset V
\end{equation}
for any $\eps\in \frac{1}{\Delta_*} N$, with the latter being a neighbourhood of $\eps_0$. Hence, by the construction (\ref{DeldotposV}) of the interval $V$, every point $\eps \in \frac{1}{\Delta_*} N$ is also a regular value of the fidelity parameter. Since $\eps_0\in \cE$ was arbitrary,  this proves that the set $\cE$ in (\ref{cE}) is open. Moreover, due to (\ref{locinv}), the differentiation of $\tau$ reduces to that of the local inverse $(\Delta|_V)^{-1}$ and, by the Inverse Function Theorem,  leads to the first equality in (\ref{tau'}), where the second equality is obtained from the first one by repeated differentiation of $\tau'$ as a composite function: $  \tau''(\eps)
  =
-
    \Delta_*\ddot{\Delta}(\tau(\eps)) \tau'(\eps)/
    \dot{\Delta}(\tau(\eps))^2
$, thus   extending (\ref{tau'''0}).
\hfill$\blacksquare$
\end{pf}

The quantum system being considered can result as an augmented system from an interconnection of component systems through field-mediated or direct energy coupling and  subjected to external quantum fields. If the system has a finite number of adjustable energy and coupling parameters which, without loss of generality, can be assembled into a  vector $p\in \mR^r$,  they can be varied so as to improve the performance of the system as a quantum memory described by the mean-square deviation functional  (\ref{Del}) and the related quantity in (\ref{Delmax}) which acquire   $p$-dependence: $\Delta=\Delta(t,p)$ and
\begin{equation}
\label{DelTp}
      \Delta_T(p)
    :=
    \max_{0 \< t \< T}
    \Delta(t,p).
\end{equation}
The minimisation of the latter yields a finite-horizon  memory optimisation  problem
\begin{equation}
\label{DelT}
    \Delta_T(p)
    \to \inf,
    \qquad
    p \in \mR^r
\end{equation}
at a given $T>0$.  An alternative (and, in a sense,   dual)   memory optimisation  problem
\begin{equation}
\label{taumax}
  \tau(\eps,p) \to \sup,
  \qquad
  p \in \mR^r
\end{equation}
(for a given $\eps>0$)
involves the corresponding $p$-dependent decoherence time (\ref{tau}):
\begin{align}
\nonumber
    \tau(\eps,p)
    & :=
    \sup
    \{
        T >0:\
        \Delta_T(p) < \eps\Delta_*
    \}\\
\label{taup}
    & =
    \min
    \{
        t\> 0:\
        \Delta(t,p)
        \>
        \eps
        \Delta_*
    \},
\end{align}
where the reference scale parameter $\Delta_*$ from (\ref{Del*}) does not depend on $p$.  Accordingly, the open set (\ref{cE}) of regular fidelity levels is also $p$-dependent:
\begin{equation}
\label{cEp}
  \cE(p)
  :=
  \{
        \eps \>0:\
        \dot{\Delta}(\tau(\eps,p), p) >0
  \},
\end{equation}
where $\dot{(\ )} := \d_t$ is now the partial derivative with respect to time $t$.

\begin{lem}
\label{lem:tau'p}
Suppose $\Delta(t,p)$ is twice continuously differentiable as a function of time $t\> 0$ and the vector $p \in \mR^r$ of system parameters, that is, $\Delta \in C^2(\mR_+\x \mR^r, \mR_+)$. Then the set-valued map $p\mapsto \cE(p)$ in (\ref{cEp})  is lower semi-continuous. Furthermore, the decoherence time $\tau(\eps,p)$ in (\ref{taup}) is twice continuously differentiable not only with respect to $\eps$ as in (\ref{tau'}), but also as a function of $(\eps, p)$ at any regular value of $\eps$, and its gradient vector over $p$, the vector of mixed second-order partial derivatives over $\eps$ and $p$,  and the  Hessian matrix over  $p$ are computed as
\begin{align}
\label{tau'p}
  \d_p \tau
  & =
  -
  \frac{1}{\dot{\Delta}}
  \d_p \Delta,\\
\label{tau'epsp}
  \d_p \tau'
  & =
  -
  \frac{1}{\dot{\Delta}}
  (\ddot{\Delta} \d_p\tau + \d_p \dot{\Delta})\tau',\\
\label{tau''p}
  \d_p^2 \tau
  & =
  -
  \frac{1}{\dot{\Delta}}
  (\ddot{\Delta} \d_p \tau \d_p \tau^\rT
  + 2\bS(\d_p\tau \d_p\dot{\Delta}^\rT)
  +
  \d_p^2 \Delta)
\end{align}
for any   $\eps \in \cE(p)$. Here, $\bS(\mu):= \frac{1}{2}(\mu + \mu^\rT)$ is the symmetrizer of square matrices, and the partial derivatives $\dot{(\ )}$ and $\d_p$ of $\Delta(t,p)$ are evaluated at $t = \tau(\eps,p)$. \hfill$\square$
\end{lem}
\begin{pf}
Since $\Delta \in C^2(\mR_+\x \mR^r, \mR_+$), then
for any given $p_0\in \mR^r$ and any given regular fidelity level $\eps_0\in \cE(p_0)$, that is, satisfying $\dot{\Delta}(T,p_0)>0$, with $T:= \tau(\eps, p_0)$,  there exist sufficiently small $\delta>0$ and $u>0$  such that
(\ref{DeldotposV}) holds for all $p$ from the open ball $B_u:= \{p \in \mR^r: |p-p_0|< u\}$  of radius $u$ and centre $p_0$, that is,
\begin{equation}
\label{DeldotposVp}
    \dot{\Delta}(t,p)>0,
    \qquad
    t \in (T-\delta, T+\delta)=:V,
    \quad
    p \in B_u,
\end{equation}
and  the equation $\Delta(t,p) = \eps_0\Delta_*$ has a unique solution $t = \phi(p) \in V$ for any $p \in B_u$,   which,  by the Implicit Function Theorem, is a smooth  function satisfying
\begin{equation}
\label{phidiff}
    (\dot{\Delta}(t,p)\d_p \phi(p)+ \d_p \Delta(t,p))|_{t = \phi(p)} = 0,
    \qquad
    p \in B_u.
\end{equation}
Moreover, since $\Delta$ is uniformly continuous on any compact subset of $\mR_+ \x \mR^r$, there exists $v \in (0,u)$ such that  (\ref{<<}) holds in a  stronger form
\begin{align}
\nonumber
    \max_{p \in \overline{B}_v}
    \Delta(T-\delta, p)
    & \<
    \max_{p \in \overline{B}_v}
    \Delta_{T-\delta}(p) =: d_1   \\
\label{<<<}
    & <
    \eps_0 \Delta_*
    <
    \min_{p \in \overline{B}_v}
    \Delta(T+\delta, p) =:d_2,
\end{align}
where use is made of (\ref{DelTp}), and
$\overline{B}_v$ is the closed ball of radius $v$ and centre $p_0$. This allows the function $\phi$ to be extended   to a function  $\psi: D\x B_v \to V$ (so that $\psi(\eps_0, p) = \phi(p)$ for any $p \in B_v$) which describes a unique  solution $t = \psi(\eps, p)$  of the equation $\Delta(t,p) = \eps\Delta_*$ with respect to $t \in V$ for any $(\eps, p) \in D\x B_v$, where $D:= \frac{1}{\Delta_*}(d_1, d_2) \ni \eps_0$.  By the construction of $d_1$, $d_2$ in (\ref{<<<}), the decoherence time (\ref{taup}) on the open set $D \x B_v$ coincides with the function $\psi$:
\begin{equation}
\label{taupsi}
    \tau|_{D\x B_v} = \psi.
\end{equation}
Moreover, with $t_1:= \max_{p \in \overline{B}_v} \tau(d_1,p)< t_2:=\min_{p \in \overline{B}_v}\tau(d_2,p)$ defined in terms of (\ref{<<<}),  the function $\Delta$ coincides with (\ref{DelTp}) in a smaller neighbourhood of $(T,p_0)$:
\begin{equation}
\label{DelDel}
  \Delta(t,p) = \Delta_t(p),
  \qquad
  t \in (t_1, t_2),
  \quad
  p \in B_v.
\end{equation}
Now, in view of (\ref{DeldotposVp}), the relation (\ref{taupsi}) implies that $D \subset \bigcap_{p \in B_v} \cE(p)$, which (due to arbitrariness of $p_0 \in \mR^r$ and  $\eps_0 \in \cE(p_0)$) establishes the lower semi-continuity $\cE(p_0)\subset \liminf_{p\to p_0} \cE(p) :=  \bigcup_{u >0}\bigcap_{p\in B_u} \cE(p)$ of the set-valued map $\cE$. The relations (\ref{tau'p})--(\ref{tau''p}) are inherited by $\tau$ from the function $\psi$ and are obtained, similarly to (\ref{phidiff}), by differentiating both sides of the equality $\Delta(\tau(\eps,p),p) = \eps \Delta_*$ (whose left-hand side is a composite function of $\eps$ and $p$) as justified by the Implicit Function Theorem.
\hfill$\blacksquare$
\end{pf}

The following theorem relates local solutions of the memory decoherence time maximisation problem (\ref{taumax}) and the finite-horizon mean-square deviation functional minimisation problem (\ref{DelT}).

\begin{thm}
\label{th:taumaxdevmin}
Suppose $\eps >0$ is a given fidelity level. If $p \in \mR^r$ delivers  a strong local maximum for the problem (\ref{taumax}), and $\eps$ is regular in the sense of (\ref{cEp}), then $p$  also delivers a strong local minimum for the problem (\ref{DelT}) at the time horizon $T = \tau(\eps,p)$.\hfill$\square$
\end{thm}
\begin{pf}
The property that $p_0 \in \mR^r$ delivers a strong local maximum in (\ref{taumax}) is equivalent to $\d_p \tau(\eps,p_0) = 0$ and $\d_p^2 \tau(\eps, p_0) \prec 0$. Under the regularity condition $\eps \in \cE(p_0)$, this implies that $\d_p \Delta(T,p_0) = 0$ and $\d_p^2 \Delta(T,p_0) \succ 0$ (evaluated at $T:= \tau(\eps,p_0)$) in view of (\ref{tau'p}),  (\ref{tau''p}) of Lemma~\ref{lem:tau'p}. Hence,  $p_0$ also delivers a strong local minimum in $\Delta(T,p)\to \inf$ over $p \in \mR^r$. The latter problem is identical to (\ref{DelT}) in a sufficiently small neighbourhood of $p_0$ in view of  the relation (\ref{DelDel}) from the proof of the lemma.
\hfill$\blacksquare$
\end{pf}

\section{Discounted memory performance criteria}
\label{app:disc}

Regardless of a particular structure of the deviation measure $\Delta$ in (\ref{Del}) (or its Schr\"{o}dinger type counterpart $\Gamma$ in  (\ref{DelHS})), the memory decoherence time $\tau$ in (\ref{tau}) has the following link with discounted performance criteria [\cite{B_1965}] and their applications to coherent quantum control [\cite{VP_2021_MTNS,VP_2022_MCSS}].

\begin{thm}
\label{th:disc}
The mean-square deviation functional $\Delta$ in (\ref{Del}) and the memory decoherence time $\tau$ in (\ref{tau}),  as functions on $\mR_+$,   satisfy
\begin{equation}
\label{tauT}
    \frac{1}{T}
    \int_0^{+\infty}
    \Delta(t)
    \re^{-\frac{t}{T}}
    \rd t
    \<
    \Delta_*
    \int_0^{+\infty}
    \re^{-\frac{\tau(\eps)}{T}}
    \rd \eps
\end{equation}
for any $T>0$ (small enough to make the integrals convergent).
\hfill$\square$
\end{thm}
\begin{pf}
The left-hand side of (\ref{tauT}) can be represented as the expectation
\begin{equation}
\label{left}
    \int_0^{+\infty}
    f_T(t)
    \Delta(t)
    \rd t
    =
    \bM \Delta(\omega_T)
\end{equation}
over an auxiliary  $\mR_+$-valued classical  random variable $\omega_T$ with an exponential probability density function (PDF)
\begin{equation}
\label{fT}
    f_T(t)
    :=
    \frac{1}{T}
    \re^{-\frac{t}{T}},
    \qquad
    t\> 0,
\end{equation}
parameterised by $T>0$,
so that the tail probability distribution  of $\omega_T$ is  given by
\begin{equation}
\label{exptail}
    \bP(\omega_T \> t)
    =
    \re^{-\frac{t}{T}}.
\end{equation}
The definition (\ref{tau}) of $\tau$ in terms of $\Delta$ implies the inclusion of the random events
$
    \{\Delta(\omega_T) > \eps \Delta_*\}
    \subset
    \{\omega_T \> \tau(\eps)\}
$ for any fixed but otherwise arbitrary $\eps>0$,
and hence, by the monotonicity of probability measures,
\begin{equation}
\label{P<P}
    \bP(\Delta(\omega_T) > \eps \Delta_*)
    \<
    \bP(\omega_T \> \tau(\eps)).
\end{equation}
Therefore, the integral representation of the mean values of nonnegative random variables in terms of their tail distributions [\cite{S_1996}], in combination with (\ref{P<P}) and (\ref{exptail}),  leads to
\begin{align}
\nonumber
    \frac{1}{\Delta_*}
    \bM\Delta(\omega_T)
    & =
    \int_0^{+\infty}
    \bP(\Delta(\omega_T) > \eps \Delta_*)
    \rd \eps\\
\label{MDel}
    & \<
    \int_0^{+\infty}
    \bP(\omega_T \> \tau(\eps))
    \rd \eps
    =
    \int_0^{+\infty}
    \re^{-\frac{\tau(\eps)}{T}}
    \rd \eps,
\end{align}
thus establishing the inequality (\ref{tauT}) in view of (\ref{left}), (\ref{fT}).
\hfill$\blacksquare$
\end{pf}

The averaging on the left-hand side of (\ref{tauT}) corresponds to considering the quantum system as a dynamic random access memory, with $\omega_T$ in the proof of Theorem~\ref{th:disc} being interpreted as an exponentially distributed  random time (with the mean value $T = \bM \omega_T$)  between storage requests forming a homogeneous Poisson process of intensity $\frac{1}{T}$. The parameter $T$ thus plays the role of an effective time horizon.

Also note that (\ref{tauT}) can be extended to integral inequalities using more general distributions, different from the exponential one. Indeed, the inequality (\ref{P<P}) and the first two relations in (\ref{MDel}) remain valid regardless of a particular distribution   of the random variable $\omega_T$.
However, it is the use of the exponential PDF (\ref{fT}) in (\ref{tauT}) that relates (\ref{tau}) to the discounted control criteria mentioned above. In what follows, for the sake of brevity,  we will use the notation
\begin{equation}
\label{MTDel}
  \bM_T \Delta
  :=
  \bM\Delta(\omega_T)
  =
  \int_0^{+\infty}
  f_T(t)
  \Delta(t)
  \rd t
\end{equation}
for the left-hand side of (\ref{tauT}) or (\ref{left}), which,
up to a factor of $\frac{1}{T}$, is the Laplace transform (evaluated at $\frac{1}{T}$) of the function $\Delta$ (the operation $\bM_T(\cdot)$ extends to matrix- and operator-valued functions on $\mR_+$).  Therefore, the following identities are inherited from the Laplace transform:
\begin{align}
\nonumber
    \bM_T\Delta
    & =
    \overbrace{\Delta(0)}^0 + T\bM_T\dot{\Delta}
    =
    \dot{\Delta}(0)T
    +T^2\bM_T\ddot{\Delta}\\
\label{MDelser}
    & =
    \sum_{k=1}^{N-1}
    \Delta^{(k)}(0)T^k
    +
    T^N
    \bM_T\Delta^{(N)},
    \qquad
    N \> 2,
\end{align}
where $\Delta^{(k)}$ is the $k$th time derivative of $\Delta$.
This plays a role in making the inequality (\ref{tauT}) asymptotically accurate in the short-horizon limit as discussed below.

\begin{lem}
\label{lem:MTDel}
Suppose the function $\Delta$ in (\ref{Del}) satisfies (\ref{Deldotpos0}), and its second time derivative has a  subexponential growth at infinity:
\begin{equation}
\label{subexp}
\limsup_{t\to +\infty}
    \frac{\ln |\ddot{\Delta}(t)|}{t}
    < +\infty.
\end{equation}
Then both sides of (\ref{tauT}) are asymptotically equivalent in the sense that
\begin{equation}
\label{MTDelasy}
    \bM_T\Delta
    \sim
    \dot{\Delta}(0)T
    \sim
    \Delta_*
    \int_0^{+\infty}
    \re^{-\frac{\tau(\eps)}{T}}
    \rd \eps,
    \qquad
    {\rm as}\
    T\to 0+.
\end{equation}
\hfill$\square$
\end{lem}
\begin{pf}
Since $\Delta$ is smooth, the condition (\ref{subexp}) is equivalent to an upper bound
\begin{equation}
\label{upper}
    |\ddot{\Delta}(t)| \< C \re^{\lambda t},
    \qquad
    t\> 0,
\end{equation}
with some positive constants $C$, $\lambda$. Its combination with the Lebesgue Dominated Convergence Theorem yields
$
    |\bM_T \ddot{\Delta}|
    \<
    \bM_T |\ddot{\Delta}|
    \<
    \frac{C}{T}
    \int_0^{+\infty}
    \re^{\lambda t - \frac{t}{T}}
    \rd t
    =
    C
    \int_0^{+\infty}
    \re^{-(1-\lambda T)v }
    \rd v
    \to C$,
as $T \to 0+$. Hence, in view of (\ref{MDelser}),  $\bM_T\Delta = \dot{\Delta}(0)T + O(T^2)$, as $T \to 0+$, thus establishing the first asymptotic equivalence in (\ref{MTDelasy}).  By repeatedly integrating (\ref{upper}) and using the relation $\Delta(0)=0$, it follows  that
$    \Delta(t)
      \<
     \dot{\Delta}(0) t + \frac{C}{\lambda} (\frac{1}{\lambda}(\re^{\lambda t} -1)-t)\< \frac{D}{\lambda} (\re^{\lambda t}-1)$ for all $t \> 0$,
where $D:= \max(\dot{\Delta}(0), \frac{C}{\lambda})>0$ is another constant. Hence, the decoherence   time $\tau$  in (\ref{tau}) admits a lower bound
\begin{equation}
\label{taulower}
    \tau(\eps)
    \>
    \frac{1}{\lambda}
    \ln
    \Big(
        1 + \frac{\lambda \Delta_* \eps}{D}
    \Big),
    \qquad
    \eps > 0.
\end{equation}
Since $\tau'(0)$ in (\ref{tau'''0}) exists due to (\ref{Deldotpos0}), then by rescaling the integration variable on the right-hand side of (\ref{tauT}) as $\eps:= \eps_T v$, it follows that
\begin{equation}
\label{right}
        \Delta_*
    \int_0^{+\infty}
    \re^{-\frac{\tau(\eps)}{T}}
    \rd \eps
    =
    \dot{\Delta}(0) T
    \int_0^{+\infty}
    \re^{-\frac{\tau(\eps_T v)}{T}}
    \rd v,
    \ \ \
    \eps_T:= \frac{T}{\tau'(0)}. \!\!\!\!
\end{equation}
In view of (\ref{tauquad}) and the Lebesque Dominated Convergence Theorem, for any $\delta>0$,
\begin{equation}
\label{lead}
    \lim_{T\to 0+}
        \int_0^\delta
    \re^{-\frac{\tau(\eps_T v)}{T}}
    \rd v
    =
    \int_0^\delta
    \re^{-v}
    \rd v.
\end{equation}
By (\ref{taulower}), the remainder of the integral in (\ref{right}) admits an upper bound
\begin{align}
\nonumber
    \int_\delta^{+\infty}
    \re^{-\frac{\tau(\eps_T v)}{T}}
    \rd v
    & \<
    \int_\delta^{+\infty}
    (
        1 + \lambda T Ev
    )^{-\frac{1}{\lambda T}}
    \rd v\\
\nonumber
    & =
    \frac{1}{E (1-\lambda T)}
        (
        1 + \lambda T E\delta
    )^{1-\frac{1}{\lambda T}}\\
\label{rest}
    & \to
    \frac{1}{E}\re^{-E \delta},
    \qquad
    {\rm as}\
    T\to 0+
\end{align}
(where the equality is valid for $\lambda T<1$),
with $E := \frac{\dot{\Delta}(0)}{D}>0$, and thus can be made arbitrarily small  by choosing $\delta$ large enough. A combination of (\ref{lead}) with (\ref{rest}) implies that
$
        \lim_{T\to 0+}
        \int_0^{+\infty}
    \re^{-\frac{\tau(\eps_T v)}{T}}
    \rd v
    =
    \int_0^{+\infty}
    \re^{-v}
    \rd v = 1
$,
whereby (\ref{right}) leads to the second equivalence in (\ref{MTDelasy}), thus completing the proof.
\hfill$\blacksquare$
\end{pf}

As demonstrated in Theorem~\ref{th:Deldisc} below,  the discounted version of the mean-square deviation functional $\Delta$ in  (\ref{Del}) can be calculated similarly to [\cite{VP_2021_MTNS,VP_2022_MCSS}] for OQHOs. For this class of quantum systems, the QSDE (\ref{dX}) takes the form
\begin{equation}
\label{OQHO}
  \rd X = A X \rd t + B \rd W,
\end{equation}
where $A \in \mR^{n\x n}$ and $B \in \mR^{n\x m}$ are constant matrices parameterised as
\begin{equation}
\label{AB}
  A= 2\Theta (R + M^\rT J M),
  \qquad
  B = 2\Theta M^\rT
\end{equation}
by the CCR matrix $\Theta := \frac{1}{2}I_{n/2}\ox \bJ$  of the quantum position-momentum system variables $X_1, \ldots, X_n$  (so that $n$ is even,  and  $[X,X^\rT] = 2i\Theta$), an energy matrix $R = R^\rT \in \mR^{n\x n}$ and a coupling matrix $M \in \mR^{m\x n}$ which specify the system  Hamiltonian $H = \frac{1}{2} X^\rT R X$ and the vector $L = MX$ of system-field coupling operators. Here, use is also made of the matrices $J$ and  $\bJ$ from (\ref{Omega}).

\begin{thm}
\label{th:Deldisc}
Suppose the effective time horizon $T>0$ in (\ref{fT}) satisfies
\begin{equation}
\label{Tsmall}
  T < \frac{1}{2\max(0, \ln \br(\re^A))},
\end{equation}
where $\br(\cdot)$ is the spectral radius of a square matrix, and  $A$ is the dynamics matrix of the OQHO (\ref{OQHO}) given by  (\ref{AB}).
Then the discounted average (\ref{MTDel}) of the function $\Delta$ from (\ref{Del}) can be computed as
\begin{equation}
\label{Del0disc}
    \bM_T \Delta
  =
  \bra
    \Sigma,
    P_T + P
    -2\bS((I_n-TA)^{-1}P)
  \ket_\rF,
\end{equation}
where $\Sigma$ is given by (\ref{FF}), and  $0 \preccurlyeq P_T=P_T^\rT \in \mR^{n\x n}$ is a unique solution of the algebraic Lyapunov equation (ALE)
\begin{equation}
\label{PTALE}
  A_T P_T + P_TA_T^\rT + \frac{1}{T} P + BB^\rT = 0,
\end{equation}
with a Hurwitz matrix
\begin{equation}
\label{AT}
  A_T := A - \frac{1}{2T} I_n,
\end{equation}
and the matrix $P$ is given by (\ref{P}).
\hfill$\square$
\end{thm}
\begin{pf}
As obtained in [\cite{VP_2024_ANZCC}], the mean-square deviation functional (\ref{Del}) for the OQHO (\ref{OQHO}) takes the form
\begin{equation}
\label{DelOQHO}
  \Delta(t)
  =
  \|F \alpha(t) \sqrt{P}\|_\rF^2
  +
  \bra
    \Sigma,
    G(t)
  \ket_\rF,
  \qquad
  t \> 0,
\end{equation}
where
\begin{equation}
\label{exp}
  \alpha(t)
  := \re^{t A} - I_n
\end{equation}
is an  auxiliary $\mR^{n\x n}$-valued function of time satisfying $\alpha(0) = 0$,
and
\begin{equation}
\label{G}
  G(t):= \int_0^t \re^{v A} BB^\rT \re^{vA^\rT} \rd v
\end{equation}
is the finite-horizon controllability Gramian of the pair $(A,B)$ governed by the Lyapunov ODE
\begin{equation}
\label{Gdot}
  \dot{G} = A G + GA^\rT + BB^\rT
\end{equation}
with the initial condition $G(0) = 0$.
By (\ref{exp}), (\ref{G}), the leading Lyapunov exponent for (\ref{DelOQHO}) admits an upper bound
$    \limsup_{t\to +\infty}
    \frac{\ln \Delta(t)}{t}
    \<
    2\max_{1\< k\< n}
    \Re \lambda_k
    =
    2\ln \br (\re^A)
$, since so do the functions $\|\alpha(t) P \alpha(t)^\rT\|_\rF$ and $\|G(t)\|_\rF$,
where $\lambda_1, \ldots, \lambda_n$ are the eigenvalues of the matrix $A$. Hence, the discounted average
\begin{equation}
\label{Del1}
    \bM_T \Delta
    =
    \bra
        \Sigma,
        \bM_T(
        \alpha P \alpha^\rT) + G_T
    \ket_\rF
\end{equation}
is well-defined for any $T>0$ satisfying the condition
$
    \frac{1}{T} > 2 \ln \br (\re^A)
$
which is equivalent to (\ref{Tsmall}). For any such $T$, the matrix
\begin{equation}
\label{GT}
    G_T
    :=
    \bM_T G
    =
    \frac{1}{T}
    \int_0^{+\infty}
    \re^{-\frac{t}{T}}
    G(t)
    \rd t
\end{equation}
satisfies $
    \bM_T \dot{G}
    =
    \frac{1}{T} G_T
$ in accordance with (\ref{MDelser}).
Hence, the discounted averaging of both sides of the ODE (\ref{Gdot}) yields
$
    \frac{1}{T} G_T
     = AG_T + G_T A^\rT + BB^\rT
$,
which is equivalent to the ALE
\begin{equation}
\label{GTALE}
  A_TG_T + G_T A_T^\rT + BB^\rT = 0,
\end{equation}
where the matrix $A_T$ in (\ref{AT}) is Hurwitz due to (\ref{Tsmall}).  In a similar fashion,
\begin{align}
\nonumber
    \bM_T(\alpha P \alpha^\rT)
     & =
    \frac{1}{T}
    \int_0^{+\infty}
    \re^{-\frac{t}{T}}
    (\re^{tA}P\re^{tA^\rT}
    -2\bS(\re^{tA} P))
    \rd t
    +P\\
\nonumber
    & =
    \frac{1}{T}
    \int_0^{+\infty}
    (\re^{tA_T}P\re^{tA_T^\rT}
    -2\bS(\re^{tA_{T/2}} P))
    \rd t
    +P\\
\label{int}
    & =
    N_T
    +P
    - 2\bS((I_n-TA)^{-1}P),
\end{align}
where (\ref{exp}) is used along with  the matrix $A_{T/2} = A-\frac{1}{T} I_n = A_T - \frac{1}{2T} I_n$ (which is also Hurwitz)   and the relation
$
    \frac{1}{T}
    \int_0^{+\infty}
    \re^{tA_{T/2}}
    \rd t
    =
    (I_n - TA)^{-1}
$.  The matrix $N_T:=     \frac{1}{T}
    \int_0^{+\infty}
    \re^{tA_T}P \re^{tA_T^\rT} \rd t$ in (\ref{int}) is a unique solution of the ALE
\begin{equation}
\label{NTALE}
  A_TN_T + N_T A_T^\rT + \frac{1}{T} P  = 0.
\end{equation}
By taking the sum of (\ref{NTALE}) and (\ref{GTALE}), it follows that the matrix
\begin{equation}
\label{PT}
    P_T
    :=
    N_T +
    G_T,
\end{equation}
which involves (\ref{GT}),
is a unique solution of the ALE (\ref{PTALE}). A combination of (\ref{int}) with (\ref{PT}) yields
$
    \bM_T(\alpha P \alpha^\rT)
     + G_T
    =
    P_T
    +P
    - 2\bS((I_n-TA)^{-1}P)
$,
whose substitution into the right-hand side of (\ref{Del1}) establishes (\ref{Del0disc}).
\hfill$\blacksquare$
\end{pf}

Due to the effective computability (described by Theorem~\ref{th:Deldisc}),   the discounted mean-square deviation functional (\ref{Del0disc}) and its minimisation (similar to that in [\cite{VP_2021_MTNS,VP_2022_MCSS}]) provide a less conservative alternative to the maximisation of the memory decoherence time (\ref{tau}) for OQHOs (\ref{OQHO}).  A similar discounted mean-square approach to memory optimisation can also be developed for finite-level open quantum systems. This possibility (to be discussed elsewhere) is based on the fact that, despite having dynamic variables of Pauli matrix [\cite{S_1994}]  type (on a finite-dimensional initial  space $\fH_0$) with a qualitatively different algebraic structure  and being governed by quasi-linear (rather than linear) QSDEs, they also have tractable moment dynamics [\cite{VP_2022_SIAM}].

Furthermore, with an appropriate modification, Theorem~\ref{th:disc} and Lemma~\ref{lem:MTDel} are  applicable to the memory decoherence time associated with the   Schr\"{o}dinger type mean-square deviation functional $\Gamma$ in (\ref{DelHS}) instead of $\Delta$. This setting benefits from computability of the Laplace transform of $\Gamma$ due to the linear evolution (\ref{sigdot}) of the reduced system state  and applies to the general QSDE (\ref{dX})  regardless  of whether it describes an OQHO or a finite-level system. To this end, the following theorem employs the adjoint of the superoperator $\cL$ from (\ref{sigdot}) (in the sense of the Hilbert-Schmidt inner product of operators on the initial system space $\fH_0$) given by
\begin{equation}
\label{cL+}
    \cL^\dagger(\sigma)
    =
    i[H_0,\sigma]
    +
    (L_0\sigma)^\rT \Omega L_0 - \frac{1}{2}\{L_0^\rT \Omega L_0, \sigma\}.
\end{equation}

\begin{thm}
\label{th:DelHSdisc}
Suppose the effective time horizon $T>0$ in (\ref{fT}) satisfies
\begin{equation}
\label{TsmallHS}
  T < \frac{1}{2\max(0, \sup \Re \fS)},
\end{equation}
where $\fS$ is the spectrum of the superoperator $\cL$ from (\ref{sigdot}).
Then the discounted average (\ref{MTDel}),  applied to the function $\Gamma$ in (\ref{DelHS}),  can be computed as
\begin{equation}
\label{DelHSdisc}
    \bM_T \Gamma
    :=
  \int_0^{+\infty}
  f_T(t)
  \Gamma(t)
  \rd t
  =
      \bra
        \sigma_0,
        \cK_T(\sigma_0)
    \ket_\rHS,
\end{equation}
where
\begin{equation}
\label{cKT}
    \cK_T
    :=
    \cQ_T - (\cI + T\cL)(\cI-T\cL)^{-1},
\end{equation}
and
\begin{equation}
\label{cQT}
    \cQ_T
    :=
    \frac{1}{T}
    \int_0^{+\infty}
    \re^{t \cL_T^\dagger}
    \re^{t \cL_T}
    \rd t
\end{equation}
is a positive semi-definite self-adjoint superoperator satisfying a Lyapunov type equation \begin{equation}
\label{cQTALE}
  \cQ_T \cL_T  + \cL_T^\dagger \cQ_T  + \frac{1}{T} \cI = 0,
\end{equation}
where
\begin{equation}
\label{cLT}
  \cL_T := \cL - \frac{1}{2T} \cI,
  \qquad
  \cL_T^\dagger := \cL^\dagger - \frac{1}{2T} \cI
\end{equation}
are associated with $\cL$ and $\cL^\dagger$ from (\ref{cL+}), and $\cI$ is the identity superoperator.
\hfill$\square$
\end{thm}
\begin{pf}
From (\ref{DelHS}) and the self-adjointness of density operators, it follows that
\begin{equation}
\label{DelHS1}
    \Gamma(t) = \|\sigma_0\|_\rHS^2 - 2\bra \sigma_0, \sigma(t)\ket_\rHS  +\|\sigma(t)\|_\rHS^2,
    \qquad
    t \> 0.
\end{equation}
The discounted averaging (\ref{MTDel}) of (\ref{DelHS1}) over the PDF (\ref{fT}) yields
\begin{equation}
\label{DelHS2}
    \bM_T \Gamma
     =
     \|\sigma_0\|_\rHS^2
     -
     2\bra \sigma_0, \bM_T \sigma\ket_\rHS
     +
       \bM_T
       (\|\sigma\|_\rHS^2).
\end{equation}
Now, the linear dynamics (\ref{sigdot}) allows the reduced system state to be represented in terms of the superoperator exponential as
\begin{equation}
\label{sigexp}
    \sigma(t) = \re^{t \cL}(\sigma_0),
    \qquad
    t \> 0.
\end{equation}
Hence, for any $T>0$ satisfying (\ref{TsmallHS}), the first discounted average on the right-hand side of (\ref{DelHS2}) takes the form of a convergent integral
\begin{align}
\nonumber
  \bM_T \sigma
  & =
  \frac{1}{T}
  \int_0^{+\infty}
  \re^{-\frac{t}{T}}
  \re^{t \cL}(\sigma_0)
  \rd t
  =
  \frac{1}{T}
  \int_0^{+\infty}
  \re^{t \cL_{T/2}}(\sigma_0)
  \rd t  \\
\label{MTsig}
    & =
    -(T\cL_{T/2})^{-1} = (\cI- T\cL)^{-1},
\end{align}
where $\cL_{T/2} = \cL - \frac{1}{T} \cI$ in accordance with (\ref{cLT}). Substitution of (\ref{MTsig}) into (\ref{DelHS2}) yields
\begin{align}
\nonumber
     \|\sigma_0\|_\rHS^2
     -&
     2\bra \sigma_0, \bM_T \sigma\ket_\rHS
     =
     \bra
        \sigma_0,
        (\cI -
        2(\cI- T\cL)^{-1})(\sigma_0)
     \ket_\rHS\\
\label{DelHS3}
     &=
    -
     \bra
        \sigma_0,
        ((\cI + T\cL)(\cI-T\cL)^{-1})(\sigma_0)
     \ket_\rHS.
\end{align}
The last term in (\ref{DelHS2})  can be  computed by using (\ref{sigexp}) and (\ref{cLT}) as
\begin{align}
\nonumber
       \bM_T
       (\|\sigma\|_\rHS^2)
        &=
       \frac{1}{T}
       \int_0^{+\infty}
       \re^{-\frac{1}{T}}
       \|
            \re^{t\cL}(\sigma_0)
        \|_\rHS^2
        \rd t\\
\nonumber
       & =
       \frac{1}{T}
       \int_0^{+\infty}
       \|
            \re^{t\cL_T}(\sigma_0)
        \|_\rHS^2
        \rd t\\
\nonumber
       & =
       \frac{1}{T}
       \int_0^{+\infty}
       \bra
            \sigma_0,
            \re^{t\cL_T^\dagger}(\re^{t\cL_T}(\sigma_0))
        \ket_\rHS
        \rd t        \\
\label{DelHS4}
        & =
       \bra
            \sigma_0,
            \cQ_T(\sigma_0)
        \ket_\rHS,
\end{align}
with $\cQ_T$ the superoperator from (\ref{cQT}) which is well-defined under the condition (\ref{TsmallHS}). The fact that $\cQ_T$ satisfies (\ref{cQTALE}) is established (similarly to the corresponding property of infinite-horizon observability Gramians) by integrating over $t\> 0$  both sides of the relation
$
        \dot{\cN}_T(t) = \cL_T^\dagger \cN_T(t) + \cN_T(t) \re^{t \cL_T}
$ for the time-varying positive semi-definite self-adjoint superoperator $\cN_T(t):= \re^{t \cL_T^\dagger}\re^{t \cL_T}$ and using $\cN_T(0) = \cI$ and $\lim_{t\to +\infty} \cN_T(t) = 0$ along with $\cQ_T = \frac{1}{T}\int_0^{+\infty}\cN_T(t)\rd t$ from (\ref{cQT}).  Substitution of (\ref{DelHS3}), (\ref{DelHS4}) into (\ref{DelHS2}) leads to (\ref{DelHSdisc}) in view of (\ref{cKT}).
\hfill$\blacksquare$
\end{pf}

In application to finite-level systems, the self-adjoint operators involved in Theorem~\ref{th:DelHSdisc} and its proof (including the reduced system state $\sigma$) are Hermitian matrices on a finite-dimensional initial Hibert space $\fH_0$. In this case,   vectorization allows the superoperators (such as $\cL$ and its exponential $\re^{t\cL}$) to be represented by complex matrices, thus,  in particular, making (\ref{cQTALE}) a usual ALE.

\section{Conclusion}
\label{sec:conc}

We have outlined an extension of the quantum memory optimisation approach, based on the mean-square deviation of system variables of interest from their initial conditions in the Heisenberg framework of quantum dynamics, towards a Schr\"{o}dinger picture type deviation measure. The latter is organised as the squared Hilbert-Schmidt distance between the reduced system state and the initial state to be stored. The memory decoherence time maximisation has been related to minimising the  mean-square deviation functionals at a finite time horizon. We have also discussed the less conservative discounted versions of these functionals, including their computation both in the Heisenberg and Schr\"{o}dinger picture formulations. The parameterisation of the discounted criteria by  the effective time horizon suggests the use of homotopy methods (similar to those in [\cite{MB_1985,VP_2021_MTNS}]) for numerical solution of the corresponding discounted memory optimisation problems, including their extension to finite-level quantum systems.

%
\end{document}